# The publication activity and migration trends of Ukrainian scientists in the social sciences and humanities during the first two years of the Russo-Ukrainian war

**Serhii Nazarovets**[1]

[1] Borys Grinchenko Kyiv Metropolitan University, 18/2 Bulvarno-Kudriavska Str., 04053 Kyiv, Ukraine; serhii.nazarovets@gmail.com

**Abstract**

This study analyses the publication activity and migration patterns of Ukrainian scholars in the social sciences and humanities (SSH) during the initial two years of the Russo-Ukrainian war. Focusing on scholars who published at least three papers, the study underscores the resilience of these scholars, who continued their academic endeavours within their homeland despite the conflict. The research utilizes data from the Social Sciences Citation Index (SSCI) and the Arts & Humanities Citation Index (AHCI) to illustrate their continued scientific contributions under adverse conditions. It also highlights the crucial role of international collaboration in supporting Ukrainian SSH research, emphasizing that such collaborations primarily manifest through joint research projects rather than relocation of scholars to foreign institutions.

*Keywords:* publication trends; migration trends; Russo-Ukrainian war; bibliometrics

## 1. Introduction

Conducting scientific research in Ukraine has always presented formidable challenges, characterized by slow and inconsistent reforms, underdeveloped international cooperation, inadequate funding, outdated equipment, and a notable gap between the higher education sector and industry (Gaind & Liverpool, 2023; Kozak et al., 2015; OECD, 2022; Schiermeier, 2019; Yegorov, 2018). The recent annexation of Crimea and Russia's hybrid warfare against Ukraine, which escalated into a full-scale Russian invasion in February 2022, has further compounded the difficulties faced by Ukrainian scientists. This war has inflicted significant damage on the Ukrainian and global scientific communities, manifesting in the physical destruction and occupation of scientific institutions, the depletion of research infrastructure and funding, and the loss of scientific personnel due to casualties, injuries, and forced migration (da Costa et al., 2022; Irwin, 2023; Oleksiyenko et al., 2021; Pereira et al., 2022; Stone, 2015).

The academic community has promptly responded to this significant military conflict, the largest in Europe since 1945, by universally condemning Russia's aggression against Ukraine. Numerous academic institutions worldwide have extended their support to Ukrainian scientists by offering them temporary work invitations (Gaind et al., 2022; Maryl et al., 2022; Rose et al., 2023). Additionally, scientific journals have actively engaged in publishing articles that scrutinize the adverse impacts of the Russian invasion on Ukrainian science. These articles explore overarching issues related to the organization of scientific endeavours during wartime, addressing both the institutional and individual challenges faced by researchers.

Following February 2022, the productivity of Ukrainian scientists may have been adversely affected by the ongoing war. These scientists could have tragically lost their lives, been captured, conscripted into the military, transitioned to volunteer work, encountered restricted access to equipment and materials, or simply did not maintain scientific pursuits at their accustomed level due to heightened stress and anxiety (Karatzias et al., 2023; Tsybuliak et al., 2023; Xu et al., 2023). A study by Lushchak et al. (2024) revealed that forced displacement from one's prior place of residence can significantly exacerbate mental health challenges stemming from the impacts of war. Consequently, even Ukrainian research migrants who have secured access to their partners' research infrastructure may confront substantial obstacles in integrating into the new academic milieu.

A significant portion of Ukrainian scientists have been compelled to emigrate and pursue employment opportunities both domestically (Suchikova et al., 2023) and abroad (Karmadonova, 2023; Kaczmarczyk, 2022; Lipták & Kincses, 2023). According to a survey conducted by de Rassenfosse et al. (2023), approximately 18.5% of Ukrainian academics have left the country. A comprehensive survey conducted by the UAScience.Reload initiative in April 2022, involving 2,173 Ukrainian scientists, revealed that the ongoing conflict has drastically affected their professional engagement and financial stability, with over 72% unable to continue their research activities as before, and a significant majority experiencing worsened economic conditions (Lutsenko et al., 2023). However, prior studies have predominantly focused on employees of research institutes and universities in general, with no efforts made to assess the potential threat posed by the departure of productive researchers in specific research fields.

In their analysis of the consequences of the Russian invasion on Ukrainian science, Ganguli & Waldinger (2023) demonstrate a significant disruption in scientific activities, including a 10% drop in research outputs and substantial damage to physical infrastructure at leading universities. The study further highlights the displacement of prominent scientists and a drastic reduction in international collaborations, particularly with Russian institutions, posing long-term challenges to the Ukrainian scientific community. However, it is important to note that the research method employed by Ganguli & Waldinger, which primarily focused on the number of publications, likely resulted in a sample that under-represents Ukrainian scholars in the social sciences and humanities.

This study examines the changes in the publication behaviour of the productive Ukrainian scientists working in SSH during the initial two years of the Russo-Ukrainian war, with a focus driven by several significant considerations. It is widely acknowledged that history and culture can profoundly influence the trajectory of SSH development (Kulczycki et al., 2018). Following Ukraine's independence in 1991, Ukrainian SSH fields were liberated from the constraints of Soviet ideology and censorship. However, the ongoing war poses a threat to the progress made in these scientific domains within Ukraine. Moreover, researchers in SSH are often more focused on local issues than in the natural sciences (Nederhof, 2006). Consequently, their research may be less dependent on direct physical participation in international scientific events. It is conceivable that Ukrainian SSH scientists have managed to sustain productive work amid the challenges of war, provided with basic physical security and minimal funding.

This study examines the subset of Ukrainian scholars in SSH who are distinguished by their prolific publication records relative to their peers. Highly productive scientists represent a small fraction of the total number of scientists in a country. Nonetheless, they play a pivotal role in shaping the research landscape and influencing the impact of research conducted within the country, both presently and in the future. This assertion is supported in studies conducted by Abramo et al. (2017), Fox & Nikivincze (2021), and Kwiek & Roszka (2023). According to Kwiek (2016), the most productive 10% of scientists contribute to 50% of scientific papers. Moreover, productive scientists not only publish their own work but also serve as mentors, sharing their knowledge and experience to help elevate other researchers within the country to international acclaim (Agrawal et al., 2017; Yin & Zhi, 2017). Therefore, the potential loss of Ukraine's productive scientists could have catastrophic consequences in the wake of the Russo-Ukrainian war. Nevertheless, the urgency of addressing the impact of the war on productive Ukrainian scientists remains unacknowledged. However, it is important to contextualize these insights within the Ukrainian scenario: the publication activity of Ukrainian scholars in leading international journals is relatively limited, and this study focuses on those who are productive within the specific confines of Ukraine's research environment. The criteria and rationale for selecting productive Ukrainian scientists in SSH are described in the Data and methods section.

The inquiry into whether there have been notable shifts in the percentage of internationally published works by the most productive Ukrainian scientists in SSH is of significant interest. Since the onset of the war, international scientists have displayed a heightened willingness to engage Ukrainian partners in academic collaboration (Oleksiyenko et al., 2023). This inclination should theoretically lay the groundwork for an uptick in the international activity of Ukrainian scientists. However, Ukrainian scientists may also encounter circumstances beyond their control that impede their ability to respond to proposals from foreign scientific partners. Furthermore, they may face resource constraints hindering the continuation of ongoing international research or find themselves compelled to redirect the focus of their research (Fiialka, 2022). Damaševičius and Zailskaitė-Jakštė (2023) observed a trend suggesting a reduction in the diversity of international collaborations among Ukrainian researchers in 2022. Nonetheless, it is crucial to note that these findings were derived from data sourced from a single academic publisher.

Another crucial aspect is to evaluate the significance of international scientific collaborations in the work and migration dynamics of Ukrainian scientists (Korennoy & Khorevin, 1995). It can be inferred that if a Ukrainian scientist had been actively engaged in collaborations with foreign counterparts prior to the onset of the Russo-Ukrainian war, then in such instances, foreign colleagues might be more inclined to extend offers for Ukrainian scientists to work in their respective countries. Concurrently, it may have been relatively easier for Ukrainian scientists to contemplate relocating to other countries under such circumstances.

It is significant to investigate the migration patterns of productive Ukrainian scientists during the war, with particular consideration of gender and age dynamics. Notably, women in Ukrainian science demonstrate higher publication productivity compared to men, a trend often observed in developing countries with limited scientific output (Larivière et al., 2013). Moreover, Ukrainian legislation restricts men aged between 18 and 60 who are mandated to serve in the military are restricted from travelling abroad during the enforcement of martial law in Ukraine[1]. Consequently, it is anticipated that female scientists will exhibit a higher migration rate than their male counterparts. However, the impact of this phenomenon on the productivity of Ukrainian scientists remains unclear.

Under martial law, Ukrainian men require a special permit even for short-term visits abroad, which makes it difficult for them to participate in international conferences and other scientific events, with potentially negative consequences for scientific productivity. However, Ukrainian scientists have been granted a deferral from mobilization, affording them the choice between military service and the continuation of their research. Consequently, despite encountering several adverse factors, the publishing activity of male scientists may not have been significantly impacted at a national level.

This study endeavours to answer two pivotal questions about the impact of the Russo-Ukrainian war on the academic sector within the social sciences and humanities:

Q1: Has the Russo-Ukrainian war affected the total number of the most productive Ukrainian scientists in SSH, who continue to operate within the country? This question also explores the age and gender characteristics of these scientists and examines whether there have been changes in the structure of their collaborations with foreign co-authors since the war began.

Q2: What is the extent of emigration among these most productive Ukrainian scientists since the onset of the war? Moreover, how many of them are currently engaged in research activities at foreign institutions?

This paper is structured as follows: The "Data and Methods" section provides detailed information on the sources and characteristics of the dataset used in this study, along with an explanation of the rationale for selecting specific sampling criteria. The "Results" section concisely presents the key findings. Finally, the "Discussion and Conclusions" section summarizes the study’s insights and discusses its limitations.

## 2. Data and methods

Ukraine has long grappled with an imperfect system for evaluating academic productivity, which predominantly relies on quantitative metrics of research activity (Abramo et al., 2023). Consequently, there is a propensity among Ukrainian authors to prioritize quantity over quality, with the aim of publishing as many papers as possible in journals indexed in relevant citation databases (Hladchenko, 2022; Hladchenko, 2023). This flawed national assessment framework contributes to a surge in publications by Ukrainian authors, particularly in local journals (Hladchenko & Moed, 2021; Nazarovets, 2020; Nazarovets & Mryglod, 2023), as well as in journals that have been delisted from citation databases, often due to breaches of research ethics (Nazarovets, 2022). Additionally, there is a notable rise in excessive self-citation within these publications (Ioannidis et al., 2019; Baccini & Petrovich, 2023).

In order to minimise the inclusion of authors of non-peer-reviewed papers in the sample for this study, highly selective lists of journals, namely two WoS CC indexes – the Social

[1] https://zakon.rada.gov.ua/laws/show/57-95-%D0%BF#Text

Sciences Citation Index (SSCI) and the Arts & Humanities Citation Index (AHCI) – were utilized in the initial stage of papers and authors selection. The editors of the SSCI and AHCI adhere to stringent content selection criteria in order to ensure that these indexes encompass the most authoritative journals in the SSH[2]. While this approach, of relying on selective journal lists, has its limitations concerning the comprehensiveness of research field representation (Pölönen et al., 2021), it ensures the inclusion of journal papers that have undergone proper peer review for this analysis. Therefore, to provide a broad quantitative overview of the publication activity of Ukrainian scientists in the SSH, the first phase of the study involved searching the SSCI and AHCI for publications within the last 10 years in which at least one author indicated their affiliation with a Ukrainian institution (CU=Ukraine). This search was conducted on 26 April 2024, with only articles and reviews being considered in the search results.

For the subsequent stage of this study, the search was limited to the years 2021 (the last year preceding the war) and 2023 (the second year of the full-scale Russo-Ukrainian war). The general quantitative attributes of Ukrainian scientists papers in SSH in these two distinct years were delineated, and a comparative analysis was conducted to discern changes in the publication activity of Ukrainian authors during this period. The examination of the author composition of these papers facilitated the identification of authors with the highest publication count in a given year. In order to identify the most prolific authors in the SSH fields, this study focused on Ukrainian scholars who had published at least three articles in a single year. This threshold was established based on the observation that Ukrainian scientists typically contribute a limited number of articles to influential journals, resulting in a relatively small cohort of scholars with publication rates exceeding three papers per year. For instance, data from the SSCI and the AHCI reveal that in 2021, Ukrainian scientists produced fewer than 400 articles and reviews. In contrast, scientists from Poland, which shares a border with Ukraine and has a comparable population size, published nearly 7,000 such works in the same year. The decision to exclude Ukrainian scholars who published only one or two papers per year was informed by the objective to analyse the productivity of the most active contributors within SSH fields, because precisely scholars with higher publication rates are often having greater influence on the development of their disciplines.

Conducting quantitative studies on the international migration of researchers poses challenges due to the dearth of reliable and up-to-date data. Consequently, such studies often rely on tracking researchers' movements by monitoring changes in the affiliations indicated in their papers (Moed & Halevi, 2014; Poitras & Larivière, 2023; Robinson-Garcia et al., 2019). This method was employed in the current study as well – initially, the affiliations of the most productive Ukrainian scientists in 2021 were established, followed by an analysis to ascertain if there were any changes in their institutional affiliations in 2023. Given that many scientists may have submitted their manuscripts in 2021, their papers published in 2022 may not yet reflect updated affiliations. Therefore, a comparison of the publications of these scientists in 2021 with their publications in 2023 was deemed appropriate. By 2023, authors would have had sufficient time to indicate their current affiliations, ensuring a more accurate assessment of migration patterns and changes in institutional affiliations.

---

[2] https://clarivate.com/products/scientific-and-academic-research/research-discovery-and-workflow-solutions/webofscience-platform/web-of-science-core-collection/editorial-selection-process/editorial-selection-process/

The SSCI and the AHCI do not encompass the entirety of scientists' publication activity. In order to gain a more comprehensive understanding of the publication behaviour of the analysed authors in the relevant years, I also determined the total number of articles and reviews authored by them in all journals indexed in the Web of Science Core Collection (WoS CC). This was achieved by using the Web of Science ResearcherID of these scientists. The limited number of publications by Ukrainian scholars in these fields over the years allowed for a manual analysis of bibliographic data to ascertain gender, age, and verify information regarding the current place of work of the researcher. To accomplish this, I manually supplemented the WoS CC data with information sourced from the official websites of the institutions and the personal pages of the researchers on specialized academic platforms such as ORCID and Google Scholar Citations.

In order to verify the accuracy of the listed affiliations of the selected Ukrainian scientists – considering the possibility of outdated information or instances where scientists might have maintained a Ukrainian institutional affiliation despite residing abroad – I contacted each scientist directly. Emails were sent to confirm their presence in Ukraine in 2023 and to ensure that their research papers were genuinely authored within the country. Additionally, I invited respondents to provide qualitative insights into their experiences of conducting scientific work during the Russo-Ukrainian war. I encouraged them to share the factors that enabled and motivated them to continue their research activities under challenging circumstances. This approach was designed to gain a deeper understanding of both the personal and professional impacts of the war on their scientific work.

## 3. Results

### *3.1 Bibliometric analysis*

Between 2014 and 2023, Ukrainian scientists contributed a total of 2,504 articles and reviews to journals indexed in the SSCI and AHCI. Notably, the number of publications exhibited a consistent increase each year, despite the backdrop of the Russo-Ukrainian war (Fig. 1). Table 1[3] presents the Ukrainian scientists affiliated with Ukrainian institutions who had at least three publications in SSCI and AHCI in 2021. A total of 22 productive researchers (11 women and 11 men) were identified in 2021. It should be noted that these authors are productive scientists within the Ukrainian SSH, but not necessarily within the global SSH. It is crucial to clarify that the method outlined in this paper for selecting scientists via SSCI and AHCI was solely used to identify Ukrainian scientists who disseminate their research findings in influential scientific journals. The lists of scientists provided should not be construed as a ranking of scientists. To prevent any potential misinterpretation, the names of the scientists in Tables 1 and 2 are presented in alphabetical order.

In 2023, the publication activity of 16 of these researchers experienced a decrease, while 2 remained unchanged, and 4 researchers showed an increase compared to their output in 2021. Information regarding the research activity and current affiliation of one individual could not be ascertained. Notably, none of the identified researchers (21) ceased indicating a Ukrainian institution in their publications. Furthermore, in 2023, four researchers (3 men and 1 woman) disclosed affiliations with foreign institutions in addition to their Ukrainian affiliations. According to the Web of Science classification, Ukrainian authors demonstrated the highest productivity in the following research areas in 2021: environmental sciences and ecology (81 publications), public environmental occupational health (46), science technology (other topics) (46), and business economics (30). Collaboration with foreign partners played a pivotal role for the majority of productive Ukrainian researchers in the field of SSH. Indeed, 72% of their publications were the result of international collaboration, while national collaborations and individual works each accounted for 14%. Noteworthy international collaborations of Ukrainian scientists in 2021 were observed with Poland (84), the USA (72), England (59), Germany (53), and Italy (44).

[3] All data presented in the tables 1 and 2 are sourced from Web of Science databases. The information is non-confidential and available for public review.

In 2023, a total of 32 productive Ukrainian scientists were identified, comprising 13 women and 19 men. Table 2 presents the Ukrainian scientists affiliated with Ukrainian institutions who had at least three publications in SSCI and AHCI in 2023. Remarkably, despite the ongoing war, 25 of these scientists managed to increase their annual publication activity compared to 2021. Additionally, besides their affiliation with Ukrainian institutions, 8 authors (4 men and 4 women) indicated affiliations with foreign institutions in 2023. Concurrently, one author had previously indicated affiliations with two institutions prior to the outbreak of the war, while another author commenced indicating a Ukrainian institution during the war. The remaining five authors solely indicated Ukrainian affiliation in their articles before the war.

In 2023, the most productive Ukrainian authors were in the following research areas according to Web of Science classification: environmental sciences and ecology – 99, science technology (other topics) – 70, psychology – 51, business economics – 49, public environmental occupational health – 46. Notably, Ukrainian scientists continued to produce a significant percentage of papers through international collaborations, accounting for 71%. Additionally, there was a modest increase in national collaborations, comprising 18%, while the proportion of single-authored publications decreased to 11%. The most productive international collaborations for Ukrainian scientists in 2023 were with Poland (109), the USA (82), China (52), Germany (48), England (47), and Italy (35). Hence, throughout the initial two years of the full-scale war, the structure of Ukraine's most productive scientific collaborations remained largely unchanged. The only noteworthy alteration was the inclusion of China in these top collaborations in 2023.

In summary, this study identified a total of 47 productive Ukrainian authors in the field of SSH for the years 2021 and 2023, encompassing 20 women and 27 men. Notably, 64% of these researchers were able to maintain or enhance their annual publication activity in journals indexed in WoS CC. Furthermore, only 11 (23%) of the researchers indicated additional foreign affiliations in their papers, alongside their Ukrainian institution. Specifically, 6 men and 5 women included such foreign affiliations in their papers. Importantly, none of the analysed authors relinquished their Ukrainian affiliation. One author altered their Ukrainian affiliation, while information regarding the subsequent research career of another productive author from 2021 could not be obtained. Additionally, the majority of researchers (77%) were exclusively engaged in a single Ukrainian institution during the period from 2021 to 2023. Notably, all male scientists included in this study were under the age of 60.

*3.2 Analysis of online survey*

A total of 46 emails were dispatched to a selected group of Ukrainian scientists, requesting information about their country of residence during their research activities in 2023. A response was received from 19 scientists, representing a response rate of 41%. All but one of the responses validated the location information provided in their respective publications. Specifically, 15 respondents confirmed that they had conducted their research within Ukraine, while four respondents, though working abroad, indicated that they had maintained active scientific collaborations with an institution in Ukraine.

In his answers to questions about the peculiarities of conducting scientific research in Ukraine during the war, several respondents highlighted an increase in international collaboration. Foreign scientists and institutions have actively sought partnerships with Ukrainian researchers, often providing essential resources and invitations to participate in international projects. This surge in support not only aids in maintaining ongoing research but also enhances the global visibility of Ukrainian scientific contributions. There is a notable increase in feedback and recognition from international peers, which has boosted morale and spurred further engagement in research activities. For many, this external validation has been crucial during times of uncertainty and stress. Financial support from international bodies and the waiving of article processing charges (APCs) by journals also have alleviated some of the financial burdens on Ukrainian scientists, allowing them to focus on their research without the additional stress of funding.

Emotional resilience plays a significant role in the continuation of scientific work. Several scientists described their research as a form of sublimation – a way to distract from the harsh realities of war and to maintain mental health. The sense of duty towards students and the scientific community also emerged as a strong motivating factor, driving researchers to work with heightened determination under difficult circumstances.

The personal accounts from the scientists provide a profound insight into their resilience. One researcher stated, "My team is here, and only by my example can I demonstrate the strength of the spirit, the desire to teach students and the opportunities to develop in Ukraine." Another highlighted the extraordinary circumstances, noting, "Among the illustrative examples, I can mention the work of colleagues who have been fighting on the front lines since the first days of the war, but in some incredible way find the opportunity to work on rotation." This commitment extends to the broader mission of science, as another scientist passionately declared, "Our mission and our passion is to do science. And whether they bomb us or want to destroy us, we will defend our culture and the right to intellectual sovereignty."

## Discussion and conclusions

In this paper, through a quantitative analysis of publication activity, the aim was to ascertain whether there were any changes in the number and publication behaviour of the most productive Ukrainian scientists in SSH during the initial two years of the Russo-Ukrainian war. The findings of the study suggest that despite the enduring conflict, Ukrainian scientists, alongside officials and other scientific stakeholders, managed to sustain scientific endeavours within the country. This resilience was facilitated, at least in part, by support from foreign partners. Contrary to pessimistic projections, there was no notable decline observed in the quantity of Ukrainian publications in the field of SSH. The question remains that, if not for the war, the growth of Ukrainian research papers could have been much greater, and it is also likely that certain negative effects will become noticeable only in a few years.

Although not all productive authors identified for 2021 were able to maintain their high publication activity in subsequent years, in only one case out of 47 analysed was there no information on the author's continuing research activity. Thereby, it is evident that Ukrainian scientists in SSH persist in their research pursuits. While the numerical significance of these scholars and their scientific outputs may appear modest in the global scale, it underscores their critical importance for the future research trajectory of Ukraine. Since achieving independence, Ukraine has steadily enhanced its contribution to the global treasury of knowledge in SSH. These scientists, along with their research work within Ukrainian institutions, play a pivotal role in disseminating and imparting knowledge to forthcoming generations of Ukrainian scientists in these fields (Fisher, 2023).

Ukrainian productive scientists in SSH persist in their work within Ukrainian institutions, with nearly a quarter of them indicating additional foreign affiliations during the course of the war. Interestingly, my analysis did not uncover any significant gender disparities in this regard. Notably, all male scientists included in this study were under the age of 60, rendering them theoretically subject to border crossing restrictions, which could potentially impede the establishment of international scientific partnerships. However, Ukrainian legislation provides male scientists with a deferral from mobilization, enabling them to sustain their research activities effectively, as evidenced by the findings of this analysis. Furthermore, the example of Ukrainian female scientists, who are exempt from military service, indicates that the opportunity for border crossing has not precipitated widespread emigration abroad among this demographic.

Presently, Ukrainian government officials frequently express their aspiration to repatriate Ukrainian academics who were compelled to depart the country following February 24, 2022[4]. However, the findings of this study underscore that the most productive Ukrainian scientists in SSH have opted to remain at the institutions where they were employed prior to the Russian invasion.

It is noteworthy that approximately three-quarters of the papers authored by productive Ukrainian scientists involve collaboration with foreign researchers, underscoring the paramount importance of international collaboration and maintaining ties with foreign partners. Amid the quarantine restrictions imposed due to the COVID-19 pandemic, many research teams gained valuable experience in fostering international scientific cooperation online (Chacón-Labella et al., 2021; Rudzin et al., 2022). It is conceivable that Ukrainian scientists were able to leverage this experience effectively during the war, although empirical evidence is required to substantiate this hypothesis.

During the initial two years of the war, the structure of international research collaborations among Ukrainian scientists remained largely unchanged, suggesting that the Russo-

[4] For instance https://www.ukrinform.ua/rubric-technology/3784518-ukraina-znacno-zbilsue-vitrati-na-naukovi-rozrobki-lisovij.html

Ukrainian war did not exert a significant impact on established models and trends of international scientific cooperation in SSH. However, it is evident that Ukraine's international scientific partnerships should be expanded. Nevertheless, amidst the Russian military invasion, Ukraine faces challenges in creating the requisite conditions and allocating resources for such expansion.

The results derived from this quantitative analysis did not provide insights into which Ukrainian scientists have availed themselves of offers from foreign partners to work and train at their institutions (Rassenfosse et al., 2023; Rose et al., 2023). Given the focus of this study on productive scientists, who typically possess extensive experience within a particular research domain, it is plausible that such individuals may encounter challenges in relocating their work to another country. Factors such as potential demotion, alterations in living conditions, among others, may not align with their established career trajectories and personal or family circumstances. Consequently, offers from foreign partners may be more appealing to early-career researchers, who are typically more inclined to embrace the associated risks of relocation and are yet to establish high positions and salaries (Khan, 2021). However, it's important to note that early-career researchers, owing to their relatively fewer publications, were not included in this analysis.

Moreover, it is essential to acknowledge that this study is confined to the analysis of SSH, and it's plausible that productive Ukrainian scientists in the natural sciences may exhibit different patterns. Given that quantitative analysis alone may not fully elucidate the underlying reasons for this behaviour, conducting additional interviews with Ukrainian scientists in the future could serve as a valuable avenue to delve deeper into this issue.

This paper focuses solely on the quantitative aspects of publication activity of Ukrainian scholars in SSH. However, delving into the citation patterns of their publications could provide valuable insights into the impact of their work. Nevertheless, given the limited time elapsed for the accumulation of citations, studying the impact of works produced during the Russo-Ukrainian war remains a prospect for future research.

Furthermore, while there may be a greater number of works by Ukrainian scholars available in Web of Science, it is important to acknowledge the known limitations in the completeness of metadata. As highlighted by Mryglod and Nazarovets (2023), not all records may be accurately identified as papers authored by Ukrainian authors due to these limitations. Additionally, the presence of multiple author profiles for scientists in the Web of Science could further contribute to the dispersion and incomplete coverage of their publication activity within this study. Also, it is important to consider the indexing delay inherent in SSCI and AHCI – publications may not appear in the databases immediately after publication, which could affect the accuracy of data analysis. These factors underscore the need for cautious interpretation of the quantitative data presented in this paper.

Ukrainian writer Larysa Kosach metaphorical pondered in her poem, "Mother, does every bird fly away for the winter?". This study's findings resonate with this sentiment, revealing that the most productive Ukrainian scientists in SSH have decided to bravely face and survive the "winter" in their homeland. Despite the slow growth in the publication activity of the Ukrainian SSH community, what remains paramount is the unyielding continuity of the scientific process amidst the backdrop of the Russo-Ukrainian war. This perseverance owes much to international cooperation and the active engagement of foreign research partners. For productive Ukrainian authors, this support manifests not merely in the form of open positions in foreign institutions, but rather through participation in collaborative and influential research projects. As the war persists, it becomes increasingly evident that Ukrainian scientists will continue to rely on such support from the global academic community to sustain their scholarly pursuits.

**Acknowledgements**
The author would like to thank all Ukrainian defenders for the possibility to finalize and publish this work.

**Competing interests**
The author do not have any competing interests.

**Funding statement**
This study was not supported by any sponsor or funder.

**References**

Abramo, G., D'Angelo, C. A., & Hladchenko, M. (2023). Assessing the effects of publication requirements for professorship on research performance and publishing behaviour of Ukrainian academics. *Scientometrics*, *128*(8), 4589–4609. https://doi.org/10.1007/s11192-023-04753-y

Abramo, G., D'Angelo, C.A. & Soldatenkova, A. (2017). How long do top scientists maintain their stardom? An analysis by region, gender and discipline: evidence from Italy. *Scientometrics*, 110, 867–877. https://doi.org/10.1007/s11192-016-2193-x

Agrawal, A., McHale, J., & Oettl, A. (2017). How stars matter: Recruiting and peer effects in evolutionary biology. *Research Policy*, 46(4), 853–867. https://doi.org/10.1016/j.respol.2017.02.007

Baccini, A., & Petrovich, E. (2023). A global exploratory comparison of country self-citations 1996-2019. *PLOS ONE*, *18*(12), e0294669. https://doi.org/10.1371/journal.pone.0294669

Chacón- Labella, J., Boakye, M., Enquist, B. J., Farfan- Rios, W., Gya, R., Halbritter, A. H., et al. (2021). From a crisis to an opportunity: Eight insights for doing science in the COVID-19 era and beyond. *Ecology and Evolution*, *11*(8), 3588–3596. https://doi.org/10.1002/ece3.7026

Damaševičius, R., & Zailskaitė-Jakštė, L. (2023). The impact of a national crisis on research collaborations: a scientometric analysis of Ukrainian authors 2019–2022. *Publications*, 11(3), 42. https://doi.org/10.3390/publications11030042

de Rassenfosse, G., Murovana, T., & Uhlbach, W.-H. (2023). The effects of war on Ukrainian research. *Humanities and Social Sciences Communications*, 10(1), 856. https://doi.org/10.1057/s41599-023-02346-x

Fiialka, S. (2022). Assessment of war effects on the publishing activity and scientific interests of Ukrainian scholars. *Knowledge and Performance Management*, 6(1), 27–37. https://doi.org/10.21511/kpm.06(1).2022.03

Fisher, K. (2023). Reimagining mentorship for doctoral student success: the potential utility of transformational leadership practices. *International Journal of Leadership in Education*, 1–17. https://doi.org/10.1080/13603124.2023.2295457

Fox, M.F., Nikivincze, I. (2021). Being highly prolific in academic science: characteristics of individuals and their departments. *Higher Education,* 81, 1237–1255. https://doi.org/10.1007/s10734-020-00609-z

Gaind, N., Abbott, A., Witze, A., Gibney, E., Tollefson, J., Irwin, A., & Van Noorden, R. (2022). Seven ways the war in Ukraine is changing global science. *Nature*, 607(7919), 440–443. https://doi.org/10.1038/d41586-022-01960-0

Gaind, N., & Liverpool, L. (2023). War shattered Ukrainian science – its rebirth is now taking shape. *Nature*, *618*(7967), 900–901. https://doi.org/10.1038/d41586-023-02031-8

Ganguli, I., & Waldinger, F. (2024). War and Science in Ukraine. *Entrepreneurship and Innovation Policy and the Economy*, *3*, 165–188. https://doi.org/10.1086/727771

Hladchenko, M. (2022). Implications of Publication Requirements for the Research Output of Ukrainian Academics in Scopus in 1999–2019. *Journal of Data and Information Science*, *7*(3), 71–93. https://doi.org/10.2478/jdis-2022-0016

Hladchenko, M. (2023). Effects of doctoral publication requirements on the research output of Ukrainian academics in Scopus. *Higher Education Quarterly*, (September 2022), 1–14. https://doi.org/10.1111/hequ.12463

Hladchenko, M., & Moed, H. F. (2021). National orientation of Ukrainian journals: means-ends decoupling in a semi-peripheral state. *Scientometrics*, *126*(3), 2365–2389. https://doi.org/10.1007/s11192-020-03844-4

Ioannidis, J. P. A., Baas, J., Klavans, R., & Boyack, K. W. (2019). A standardized citation metrics author database annotated for scientific field. *PLOS Biology*, *17*(8), e3000384. https://doi.org/10.1371/journal.pbio.3000384

Irwin, A. (2023). The fight to keep Ukrainian science alive through a year of war. Nature, 614(7949), 608–612. https://doi.org/10.1038/d41586-023-00508-0

Kaczmarczyk, P. (2022). Ukrainian migrants in Poland during the war: the state of the art and key challenges. *Social Insurance. Theory and Practice*, 155(4), 1-28. https://doi.org/10.5604/01.3001.0016.2359

Karatzias, T., Shevlin, M., Ben- Ezra, M., McElroy, E., Redican, E., Vang, M. L., et al. (2023). War exposure, posttraumatic stress disorder, and complex posttraumatic stress disorder among parents living in Ukraine during the Russian war. *Acta Psychiatrica Scandinavica*, *147*(3), 276–285. https://doi.org/10.1111/acps.13529

Karmadonova, T. (2023). The long way home: Migration trends of Ukrainian researchers in the modern world (1991–2023). *History of Science and Technology*, *13*(2), 263–279. https://doi.org/10.32703/10.32703/2415-7422-2023-13-2-263-279

Khan, J. (2021). European academic brain drain: A meta‐ synthesis. *European Journal of Education*, *56*(2), 265–278. https://doi.org/10.1111/ejed.12449

Korennoy, A. A., & Khorevin, V. I. (1995). Ups and downs of life sciences in Ukraine, 1979 to 1994: scientometric analysis. *Research Evaluation*, *5*(2), 143–150. https://doi.org/10.1093/rev/5.2.143

Kozak, M., Bornmann, L., & Leydesdorff, L. (2015). How have the Eastern European countries of the former Warsaw Pact developed since 1990? A bibliometric study. *Scientometrics*, *102*(2), 1101–1117. https://doi.org/10.1007/s11192-014-1439-8

Kulczycki, E., Engels, T. C. E., Pölönen, J., Bruun, K., Dušková, M., Guns, R., et al. (2018). Publication patterns in the social sciences and humanities: evidence from eight European countries. *Scientometrics*, *116*(1), 463–486. https://doi.org/10.1007/s11192-018-2711-0

Kwiek, M. (2016). The European research elite: a cross-national study of highly productive academics in 11 countries. *Higher Education*, 71(3), 379–397. https://doi.org/10.1007/s10734-015-9910-x

Kwiek, M., Roszka, W. (2023). Once highly productive, forever highly productive? Full professors' research productivity from a longitudinal perspective. *Higher Education,* 87, 519–549. https://doi.org/10.1007/s10734-023-01022-y

Larivière, V., Ni, C., Gingras, Y., Cronin, B., & Sugimoto, C. R. (2013). Global gender disparities in science. *Nature*, *504*(7479), 211–213. https://doi.org/10.1038/504211a

Lipták, K., & Kincses, Á. (2023). International migration of Ukrainian citizens to Central Europe before the Russo-Ukrainian wars. *Regional Statistics*, 13(5), 781–796. https://doi.org/10.15196/RS130501

Lushchak, O., Velykodna, M., Bolman, S., Strilbytska, O., Berezovskyi, V., & Storey, K. B. (2024). Prevalence of stress, anxiety, and symptoms of post-traumatic stress disorder among Ukrainians after the first year of Russian invasion: a nationwide cross-sectional study. *The Lancet Regional Health - Europe*, 36, 100773. https://doi.org/10.1016/j.lanepe.2023.100773

Lutsenko, Anastasiia and Harashchenko, Nataliia and Hladchenko, Lidiia and Korytnikova, Nadezhda and Moskotina, Ruslana and Pravdyva, Oleksandra, The Results of The Survey on The Needs of Ukrainian Scientists (First Wave Report) (2023). Max Planck Institute for Innovation & Competition Research Paper No. 23-03, Available at SSRN: https://ssrn.com/abstract=4335098

Maryl, M., Jaroszewicz, M., Degtyarova, I., Polishchuk, Y., Pachocka, M., & Wnuk, M. (2022). Beyond Resilience: Professional Challenges, Preferences, and Plans of Ukrainian Researchers Abroad. Zenodo. https://doi.org/10.5281/zenodo.7380509

Moed, H.F., & Halevi, G. (2014). A bibliometric approach to tracking international scientific migration. *Scientometrics*, 101, 1987–2001. https://doi.org/10.1007/s11192-014-1307-6

Mryglod, O., & Nazarovets, S. (2023). Lost for the country: country-undefined papers in Web of Science and Scopus. *Scientometrics*, *128*(4), 2619–2622. https://doi.org/10.1007/s11192-023-04661-1

Nazarovets, S. (2020). Controversial practice of rewarding for publications in national journals. *Scientometrics*, *124*(1), 813–818. https://doi.org/10.1007/s11192-020-03485-7

Nazarovets, S. (2022). Analysis of publications by authors of Ukrainian institutes in Scopus-delisted titles. *Learned Publishing*, *35*(4), 499–515. https://doi.org/10.1002/leap.1464

Nazarovets, S., & Mryglod, O. (2023). Ukrainian arts and humanities research in Scopus: a bibliometric analysis. *Library Hi Tech*. https://doi.org/10.1108/LHT-05-2023-0180

Nederhof, A. J. (2006). Bibliometric monitoring of research performance in the Social Sciences and the Humanities: A Review. *Scientometrics*, *66*(1), 81–100. https://doi.org/10.1007/s11192-006-0007-2

OECD (2022), "The future of science in Ukraine: Actions now will affect post-war recovery", OECD Policy Responses on the Impacts of the War in Ukraine, OECD Publishing, Paris, https://doi.org/10.1787/afbd05df-en.

Oleksiyenko, A., Shchepetylnykova, I., & Furiv, U. (2023). Internationalization of higher education in tumultuous times: transformative powers and problems in embattled Ukraine. *Higher Education Research & Development*, 42(5), 1103–1118. https://doi.org/10.1080/07294360.2023.2193727

Oleksiyenko, A., Terepyshchyi, S., Gomilko, O., & Svyrydenko, D. (2021). ‘What Do You Mean, You Are a Refugee in Your Own Country?’: Displaced Scholars and Identities in Embattled Ukraine. *European Journal of Higher Education*, 11(2), 101–118. https://doi.org/10.1080/21568235.2020.1777446

Pereira, P., Zhao, W., Symochko, L., Inacio, M., Bogunovic, I., & Barcelo, D. (2022). The Russian- Ukrainian armed conflict will push back the sustainable development goals. *Geography and Sustainability*, 3(3), 277–287. https://doi.org/10.1016/j.geosus.2022.09.003

Poitras, C., & Larivière, V. (2023). Research mobility to the United States: a bibliometric analysis. *Scientometrics,* 128, 2601–2614. https://doi.org/10.1007/s11192-023-04657-x

Pölönen, J., Guns, R., Kulczycki, E., Sivertsen, G., & Engels, T. C. E. (2021). National Lists of Scholarly Publication Channels: An Overview and Recommendations for Their Construction and Maintenance. *Journal of Data and Information Science*, *6*(1), 50–86. https://doi.org/10.2478/jdis-2021-0004

Robinson-Garcia, N., Sugimoto, C. R., Murray, D., Yegros-Yegros, A., Larivière, V., and Costas, R. (2019). The many faces of mobility: Using bibliometric data to measure the movement of scientists. *Journal of Informetrics*, 13(1), 50–63. https://doi.org/10.1016/j.joi.2018.11.002

Rose, Michael and Jurikova, Katarina and Pelepets, Marina and Slivko, Olga and Yereshko, Julia, Scientific Support Offers for Ukrainians: Determinants, Reasons and Consequences (2024). Max Planck Institute for Innovation & Competition Research Paper No. 24-01, Available at SSRN: https://ssrn.com/abstract=4683969

Rudzin, J. E., Soule, D. C., Whitaker, J., Berger, H., Clayton, S., & Fogaren, K. E. (2022). Catalyzing Remote Collaboration During the COVID-19 Pandemic and Beyond: Early Career Oceanographers Adopt Hybrid Open Science Framework. *Frontiers in Marine Science*, *9*. https://doi.org/10.3389/fmars.2022.855192

Schiermeier, Q. (2019). Ukraine's science revolution stumbles five years on. *Nature*, *566*(7743), 162–163. https://doi.org/10.1038/d41586-019-00512-3

Stone, R. (2015). A science exodus from Ukraine's rebellious East. *Science*, *347*(6217), 14–17. https://doi.org/10.1126/science.347.6217.14

Suchikova, Y., Tsybuliak, N., Lopatina, H., Shevchenko, L., & I. Popov, A. (2023). Science in times of crisis: How does the war affect the efficiency of Ukrainian scientists? *Problems and Perspectives in Management*, 21(1), 408–424. https://doi.org/10.21511/ppm.21(1).2023.35

Xu, W., Pavlova, I., Chen, X., Petrytsa, P., Graf-Vlachy, L., & Zhang, S. X. (2023). Mental health symptoms and coping strategies among Ukrainians during the Russia-Ukraine war in March 2022. *International Journal of Social Psychiatry*, *69*(4), 957–966. https://doi.org/10.1177/00207640221143919

Yegorov, I. (2018). Key Tendencies of Scientific and Technological Development in Ukraine and Its International Dimension (Review of Statistical Indicators). In: Valsiner, J., Lutsenko, A., Antoniouk, A. (eds) Sustainable Futures for Higher Education. Cultural Psychology of Education, vol 7. Springer, Cham. https://doi.org/10.1007/978-3-319-96035-7_7

Yin, Z., Zhi, Q. (2017) Dancing with the academic elite: a promotion or hindrance of research production?. *Scientometrics,* 110, 17–41. https://doi.org/10.1007/s11192-016-2151-7

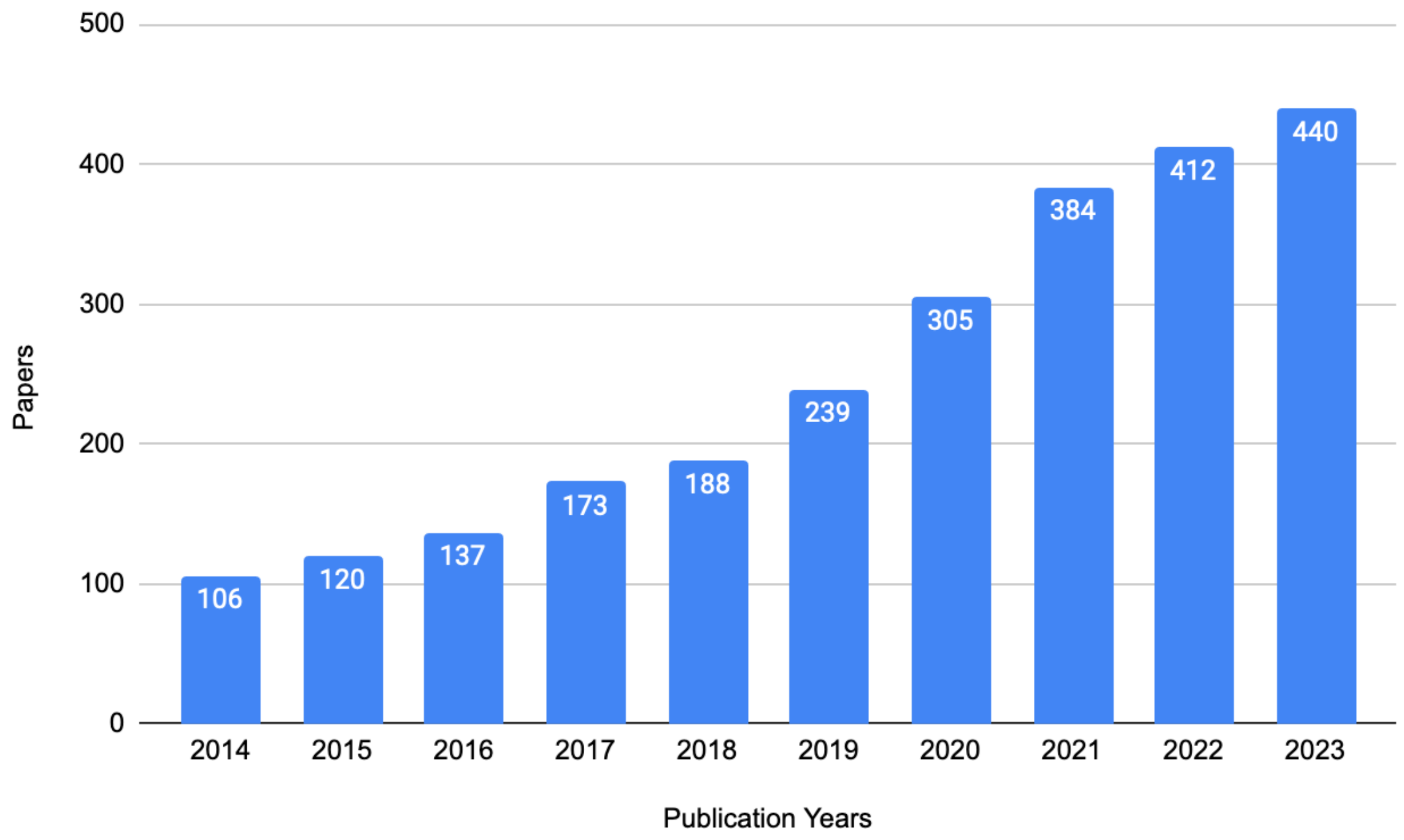

**Figure 1.** Articles and reviews of Ukrainian scientists published in 2014-2023 in the journals covered by SSCI and AHCI

**Table 1.** Authors affiliated with Ukrainian institutions with at least three publications in SSCI and AHCI in 2021

| Authors | Papers in 2021 | | Affiliation in 2021 | Gender | Papers in 2023 | | Affiliation in 2023 |
|---|---|---|---|---|---|---|---|
| | SSCI & AHCI | WoS CC | | | SSCI & AHCI | WoS CC | |
| Bogdanov, Sergiy | 3 | 6 | National University of Kyiv Mohyla Academy, Kyiv, Ukraine | Male | 0 | 2 | National University of Kyiv Mohyla Academy, Kyiv, Ukraine |
| Danyliuk, Ivan | 3 | 9 | Taras Shevchenko National University of Kyiv, Kyiv, Ukraine | Male | 5 | 8 | Taras Shevchenko National University of Kyiv, Kyiv, Ukraine |
| Dumchev, Kostyantyn | 6 | 11 | Ukrainian Institute on Public Health Policy, Kyiv, Ukraine | Male | 0 | 3 | Ukrainian Institute on Public Health Policy, Kyiv, Ukraine |
| Fedoriak, Mariia | 3 | 5 | Yuriy Fedkovych Chernivtsi National University, Chernivtsi, Ukraine | Female | 2 | 2 | Yuriy Fedkovych Chernivtsi National University, Chernivtsi, Ukraine |
| Galkin, Andrii | 3 | 6 | O. M. Beketov National University of Urban Economy in Kharkiv, Kharkiv, Ukraine | Male | 0 | 1 | O. M. Beketov National University of Urban Economy in Kharkiv, Kharkiv, Ukraine |
| Kiriazova, Tetiana | 6 | 9 | Ukrainian Institute on Public Health Policy, Kyiv, Ukraine | Female | 4 | 4 | Ukrainian Institute on Public Health Policy, Kyiv, Ukraine |
| Korolyov, Igor | 4 | 6 | Taras Shevchenko National University of Kyiv, Kyiv, Ukraine | Male | 0 | 0 | Taras Shevchenko National University of Kyiv, Kyiv, Ukraine |
| Kozytska, Inna | 4 | 4 | Taras Shevchenko National University of Kyiv, Kyiv, Ukraine | Female | 4 | 4 | Taras Shevchenko National University of Kyiv, Ukraine |
| Kurapov, Anton | 5 | 11 | Taras Shevchenko National University of Kyiv, Kyiv, Ukraine | Male | 8 | 8 | Taras Shevchenko National University of Kyiv, Kyiv, Ukraine; **University of Salzburg, Austria*** |
| Lozynskyi, Vasyl | 3 | 10 | Dnipro University of Technology, Dnipro, Ukraine | Male | 0 | 4 | Dnipro University of Technology, Dnipro, Ukraine; **Guangdong University of Petrochemical Technology, Maoming, China** |
| Lyulyov, Oleksii | 3 | 10 | Sumy State University, Sumy, Ukraine | Male | 11 | 22 | Sumy State University, Sumy, Ukraine |
| Mryglod, Olesya | 3 | 3 | Institute for Condensed Matter Physics, National Academy of Sciences of Ukraine | Female | 0 | 0 | Institute for Condensed Matter Physics, National Academy of Sciences of Ukraine |
| Nehuliaieva, Liliia | 3 | 3 | Andrei Krupynskyi Lviv Medical Academy, Lviv, Ukraine | Female | 0 | 0 | *No information* |
| Niemets, Liudmyla | 3 | 6 | V.N. Karazin Kharkiv National University, Kharkiv, Ukraine | Female | 0 | 0 | V.N. Karazin Kharkiv National University, Kharkiv, Ukraine |
| Pavlova, Daria | 3 | 3 | Ukrainian Institute for Social Research after Olexander Yaremenko, Kyiv, Ukraine | Female | 0 | 1 | Ukrainian Institute for Social Research after Olexander Yaremenko, Kyiv, Ukraine |
| Pavlova, Iuliia | 6 | 9 | Lviv State University of Physical Culture, Lviv, Ukraine | Female | 7 | 10 | Lviv State University of Physical Culture, Lviv, Ukraine |
| Pimonenko, Tetyana | 3 | 10 | Sumy State University, Sumy, Ukraine | Female | 12 | 23 | Sumy State University, Sumy, Ukraine |
| Plastun, Alex | 3 | 6 | Sumy State University, Sumy, Ukraine | Male | 0 | 1 | Sumy State University, Sumy, Ukraine |
| Saik, Pavlo | 3 | 6 | Dnipro University of Technology, Dnipro, Ukraine | Male | 2 | 4 | Dnipro University of Technology, Dnipro, Ukraine; **Guangdong University of Petrochemical Technology, Maoming, China** |
| Shevchenko, Tetiana | 3 | 6 | Sumy National Agrarian University, Sumy, Ukraine | Female | 2 | 2 | Sumy National Agrarian University, Sumy, Ukraine; **Université Paris-Saclay, Gif-sur-Yvette, France** |
| Tkachenko, Victoria | 3 | 3 | Shupyk National Healthcare University of Ukraine | Female | 0 | 0 | Shupyk National Healthcare University of Ukraine |
| Yamelynets, Taras | 3 | 4 | Ivan Franko National University, Lviv, Ukraine | Male | 0 | 0 | Ivan Franko National University, Lviv, Ukraine |

** Non-Ukrainian institutions are **bolded** in the table for clarity.*

**Table 2.** Authors affiliated with Ukrainian institutions with at least three publications in SSCI and AHCI in 2023

| Authors | Papers in 2021 | | Affiliation in 2021 | Gender | Papers in 2023 | | Affiliation in 2023 |
|---|---|---|---|---|---|---|---|
| | SSCI & AHCI | WoS CC | | | SSCI & AHCI | WoS CC | |
| Artyukhov, Artem | 0 | 7 | Sumy State University, Sumy, Ukraine | Male | 3 | 4 | **University of Economics Bratislava, Bratislava, Slovakia;** Sumy State University, Sumy, Ukraine |
| Brik, Tymofii | 0 | 2 | Kyiv School of Economics, Kyiv, Ukraine | Male | 4 | 6 | Kyiv School of Economics, Kyiv, Ukraine |
| Chumachenko, Dmytro | 0 | 1 | National Aerospace University "Kharkiv Aviation Institute", Kharkiv, Ukraine | Male | 3 | 8 | National Aerospace University "Kharkiv Aviation Institute", Kharkiv, Ukraine |
| Danyliuk, Ivan | 3 | 9 | Taras Shevchenko National University of Kyiv, Kyiv, Ukraine | Male | 5 | 7 | Taras Shevchenko National University of Kyiv, Kyiv, Ukraine |
| Diachenko, Aleksandr | 0 | 2 | Institute of Archaeology, National Academy of Sciences of Ukraine, Ukraine | Male | 4 | 6 | Institute of Archaeology, National Academy of Sciences of Ukraine, Ukraine |
| Dluhopolskyi, Oleksandr | 0 | 6 | West Ukrainian National University, Ternopil, Ukraine; Ternopil Volodymyr Hnatiuk National Pedagogical University, Ternopil, Ukraine | Male | 7 | 9 | West Ukrainian National University, Ternopil, Ukraine; **WSEI University, Lublin, Poland*** |
| Frankova, Iryna | 0 | 1 | Bogomolets National Medical University, Kyiv, Ukraine | Female | 3 | 4 | Bogomolets National Medical University, Kyiv, Ukraine; **Vrije Universiteit Amsterdam, Amsterdam, Netherlands** |
| Hutsol, Taras | 0 | 4 | Polissia National University, Zhytomyr, Ukraine | Male | 5 | 15 | Polissia National University, Zhytomyr, Ukraine |
| Kiriazova, Tetiana | 6 | 9 | Ukrainian Institute on Public Health Policy, Kyiv, Ukraine | Female | 4 | 7 | Ukrainian Institute on Public Health Policy, Kyiv, Ukraine |
| Kovalov, Bohdan | 0 | 1 | Sumy State University, Sumy, Ukraine | Male | 3 | 4 | Sumy State University, Sumy, Ukraine |
| Kozytska, Inna | 4 | 4 | Taras Shevchenko National University of Kyiv, Kyiv, Ukraine | Female | 4 | 4 | Taras Shevchenko National University of Kyiv, Ukraine |
| Krupelnytska, Liudmyla | 0 | 1 | Taras Shevchenko Natl Univ Kyiv, Kyiv, Ukraine | Female | 3 | 3 | Taras Shevchenko Natl Univ Kyiv, Kyiv, Ukraine |
| Kubatko, Oleksandr | 0 | 3 | Sumy State University, Sumy, Ukraine | Male | 3 | 3 | Sumy State University, Sumy, Ukraine |
| Kurapov, Anton | 5 | 11 | Taras Shevchenko National University of Kyiv, Kyiv, Ukraine | Male | 8 | 8 | Taras Shevchenko National University of Kyiv, Kyiv, Ukraine; **University of Salzburg, Austria** |
| Litvinova, Olena | 0 | 0 | National University of Pharmacy, Kharkiv, Ukraine | Female | 3 | 6 | National University of Pharmacy, Kharkiv, Ukraine; **Medical University of Vienna, Vienna, Austria** |
| Lyeonov, Serhiy | 0 | 1 | Sumy State University, Sumy, Ukraine | Male | 6 | 10 | Sumy State University, Sumy, Ukraine |
| Lyulyov, Oleksii | 3 | 10 | Sumy State University, Sumy, Ukraine | Male | 13 | 25 | Sumy State University, Sumy, Ukraine |
| Martsenkovskyi, Dmytro | 1 | 1 | Bogomolets National Medical University, Kyiv, Ukraine | Male | 5 | 7 | Bogomolets National Medical University, Kyiv, Ukraine |
| Meteliuk, Anna | 0 | 2 | International Charitable Foundation Alliance for Public Health, Kyiv, Ukraine | Female | 3 | 5 | International Charitable Foundation Alliance for Public Health, Kyiv, Ukraine |
| Mishchuk, Halyna | 0 | 4 | National University of Water and Environmental Engineering, Rivne, Ukraine | Female | 4 | 7 | National University of Water and Environmental Engineering, Rivne, Ukraine; **Pan-European University, Bratislava, Slovakia** |
| Nazarovets, Serhii | 0 | 1 | State Scientific & Technical Library of Ukraine, Kyiv, Ukraine | Male | 4 | 5 | Borys Grinchenko Kyiv University, Kyiv, Ukraine |
| Neyter, Roman | 0 | 0 | Kyiv School of Economics, Kyiv, Ukraine | Male | 3 | 3 | Kyiv School of Economics, Kyiv, Ukraine |
| Nosova, Bogdana | 0 | 0 | Taras Shevchenko National University of Kyiv, Kyiv, Ukraine | Female | 3 | 4 | Taras Shevchenko National University of Kyiv, Kyiv, Ukraine |
| Pavlova, Iuliia | 6 | 9 | Lviv State University of Physical Culture, Lviv, Ukraine | Female | 7 | 9 | Lviv State University of Physical Culture, Lviv, Ukraine |
| Petrytsa, Petro | 0 | 0 | Ternopil Volodymyr Hnatiuk National Pedagogical University, Ternopil, Ukraine | Male | 3 | 4 | Ternopil Volodymyr Hnatiuk National Pedagogical University, Ternopil, Ukraine |
| Pimonenko, Tetyana | 3 | 10 | Sumy State University, Sumy, Ukraine | Female | 13 | 24 | Sumy State University, Sumy, Ukraine |
| Polishchuk, Volodymyr | 0 | 2 | Uzhhorod National University, Uzhhorod, Ukraine | Male | 6 | 6 | Uzhhorod National University, Uzhhorod, Ukraine |
| Pykalo, Iryna | 1 | 1 | Ukrainian Institute of Public Health Policy, Kyiv, Ukraine | Female | 4 | 5 | Ukrainian Institute of Public Health Policy, Kyiv, Ukraine |
| Smaliychuk, Anatoliy | 0 | 1 | Ivan Franko National University of Lviv, Lviv, Ukraine | Male | 3 | 3 | Ivan Franko National University of Lviv, Lviv, Ukraine |
| Trach, Roman | 0 | 7 | **Warsaw University of Life Sciences, Warsaw, Poland** | Male | 3 | 6 | **Warsaw University of Life Sciences, Warsaw, Poland;** National University of Water and Environmental Engineering, Rivne, Ukraine |
| Trach, Yuliia | 0 | 6 | National University of Water and Environmental Engineering, Rivne, Ukraine; **Warsaw University of Life Sciences, Warsaw, Poland** | Female | 4 | 6 | National University of Water and Environmental Engineering, Rivne, Ukraine; **Warsaw University of Life Sciences, Warsaw, Poland** |
| Velykodna, Mariana | 0 | 2 | Kryvyi Rih State Pedagogical University, Kryvyi Rih, Ukraine | Female | 3 | 5 | Kryvyi Rih State Pedagogical University, Kryvyi Rih, Ukraine |

* *Non-Ukrainian institutions are* ***bolded*** *in the table for clarity.*